\documentclass[12pt]{article}
\begin{document}
\baselineskip=20pt

\title{Compact AdS space, Brane geometry and  
the AdS/CFT correspondence}

\author{\large Henrique Boschi-Filho\footnote{\noindent e-mail: 
boschi @ if.ufrj.br}\,  
and 
Nelson R. F. Braga\footnote{\noindent e-mail: braga @ if.ufrj.br}
\\ 
\\ 
\it Instituto de F\'\i sica, Universidade Federal do Rio de Janeiro\\
\it Caixa Postal 68528, 21945-970  Rio de Janeiro, RJ, Brazil}
 
\date{}

\maketitle

\vskip 3cm

\begin{abstract}
The $AdS/CFT$ correspondence can be realized in spaces that are globally different
but share the same asymptotic behavior. 
Two known cases are: a compact  $AdS$ space and the space generated by
a large number of coincident branes.
We discuss the physical consistency, in the sense of the Cauchy problem, 
of these two formulations.  
We show that the role of the boundary in the compact $AdS$ space
is equivalent to that of the flat asymptotic region in the brane space.
We also show, by introducing  a second coordinate chart for the pure $AdS$ space,
that a point at its spatial infinity corresponds to a horizon in the brane system.

\end{abstract}


\vfill\eject

\section{Introduction}
According to the Maldacena conjecture\cite{Malda} the large $N$ limit of $SU(N)$
superconformal field theories in $n$ dimensions can be described by supergravity on 
anti de Sitter ($AdS$) space-time in $n+1$ dimensions. This is known as the
$AdS/CFT$ correspondence. 
By supergravity one  means the tree level approximation of  string or M-theory
defined on $AdS_{n+1} \times M_{d}$, where $M_{d}$ is some
$d$-dimensional compactification space.
In this correspondence (see also refs. \cite{Malda2,Pe,Kle,BS} ) 
the $AdS$ space shows up both as a near horizon geometry of a set of
coincident D3-branes or as a solution of ten dimensional supergravity
(a Dirichlet $p$-brane or D$p$-brane is a $p+1$ dimensional hyperplane
where strings are allowed to end\cite{HS,Po}).

Precise prescriptions for the realization of the $AdS/CFT$ correspondence 
were presented in \cite{GKP,Wi} by considering  Poincare patches of $AdS$ space.
The Poincare coordinate system allows
a simple definition for the flat boundary where the conformal field theory 
is defined.  However there are some differences in the spaces considered in these 
references that we will discuss in this article.
Gubser, Klebanov and Polyakov \cite{GKP} started with a space generated by a 
large number $N$ of coincident D3-branes.
This space can be approximated  by an $AdS$ near the branes
and  a flat space far from them. 
On the other side, Witten \cite{Wi} has considered an $AdS$ space in Poincare 
coordinates but compactified by the inclusion of the boundary. 
These two formulations lead to equivalent results
in the sense that conformal boundary correlation functions are the same
(see also \cite{MV,FMMR}).

The approximation of the D3-brane metric as an $AdS$ space near the branes
is valid as long as the axial $AdS$ coordinate is smaller than a parameter that
increases with $N$. 
In the Maldacena conjecture the large $N$ limit is considered. 
So one might think that in this case the D3-brane space becomes a pure $AdS$ 
(without the boundary). 
However a consistent quantization is not possible in an $AdS$ space without boundary
because of  the absence of a well defined Cauchy problem\cite{QAdS1,QAdS2}.
In the formulation of \cite{GKP} the $AdS$ space is always complemented by the flat space 
asymptotic region far from the branes. We will see that this guarantees a well posed 
Cauchy problem.
In the work of \cite{Wi} there is no asymptotic flat region so
it is necessary to introduce a compactification of the $AdS$ space
for physical consistency, as we will discuss in section {\bf 3}.

Recently we have investigated the quantization of scalar fields  in the
$AdS$ bulk in terms of Poincare coordinates\cite{BB1,BB2}. 
The compactification in this coordinate system requires the introduction 
of a point at infinity which can only be properly accommodated in a second coordinate 
chart.
The two coordinate charts must match at some finite value of the
axial coordinates implying a discretization of the field spectrum.
Then it is possible to find a one to one mapping between bulk and  
boundary  quantum states, at least for scalar fields\cite{BB3}.
One can then ask: Does this extra point at infinity has any  physical role or is 
it just a mathematical tool  for a consistent quantization?
We answer this question in section {\bf 2} by constructing explicitly a 
second coordinate chart complementing the original Poincare one. 
We will see that the point at infinity represents, in the pure $AdS$
space, the horizon  that is found in the D3-brane metric. 
Curiously the complete compactification of $AdS$ space in Poincare coordinates
introduces a new horizon not present in the brane system. We also find an interpretation 
for this horizon.

\section{$AdS$ space and compactification}
 
We will start considering a pure $AdS$ space of $\,n+1$ dimensions.
This space  can be represented 
as the hyperboloid ($\Lambda\,=\,$constant)

\begin{equation}
X_0^2 + X_{n+1}^2 - \sum_{i=1}^n X_i^2\,=\,\Lambda^2
\end{equation}

\noindent in a flat $n+2$ dimensional space with measure
\begin{equation}
ds^2\,=\, - d X_0^2 - dX_{n+1}^2 + \sum_{i=1}^n dX_i^2.
\end{equation}

The  so called global coordinates $\,\rho,\tau,\Omega_i\,$ for 
$AdS_{n+1}\,$  can be defined by \cite{Malda2,Pe}

\begin{eqnarray}
\label{global}
X_0 &=& \Lambda \,\sec\rho\, \cos \tau \nonumber\\
X_i &=& \Lambda \,\tan \rho\, \,\Omega_i\,\,\,;\,\,\,\,\,\,\,
(\,\sum_{i=1}^n \,\Omega^2_i\,=\,1\,) \nonumber\\
X_{n+1} &=& \Lambda \sec \rho \,\sin\tau \,,
\end{eqnarray}

\noindent with ranges $0\le \rho <\pi/2$ and $0\le\tau< 2\pi\,$. The line element
 has the form

\begin{equation}
ds^2 \,=\, {\Lambda^2 \over cos^2 (\rho )} \Big( -d\tau^2 + d\rho^2 + 
sin^2 (\rho )d\Omega^2 \Big)\,\,.
\end{equation}

In order to identify $\tau $ as an usual time coordinate
it is necessary to unwrap it. This can be done by taking copies of the original 
$AdS$ space that together represent the $AdS$ covering space\cite{QAdS1}. 
For simplicity we will continue to call this covering space as $AdS$ as is usual
in the literature.

A consistent  quantum field theory in $AdS$ space  requires the addition of a boundary
at spatial infinity: $\rho \,=\, \pi/2\,$ in global coordinates.
This compactification of the space makes it possible to impose appropriate 
conditions and find a well defined Cauchy
problem. (Otherwise massless particles could go to or come from 
spatial infinity in finite times.) 
This result was established in \cite{QAdS1,QAdS2}.

On the other hand, $AdS$ space can be represented by Poincar\'e coordinates
$\,z \,,\,\vec x\,,\,t\,$  that are more useful for the study  of the $AdS/CFT$
correspondence. These coordinates are defined by

\begin{eqnarray}
\label{Poincare}
X_0 &=& {1\over 2z}\,\Big( \,z^2\,+\,\Lambda^2\,
+\,{\vec x}^2\,-\,t^2\,\Big)
\nonumber\\
X_j &=& {\Lambda x^j \over z}\,\,\,;\,\,\,\,\,\,\,\,\,\,\,(j=1,...,n-1)
\nonumber\\
X_n &=& - {1\over 2z}\,
\Big( \,z^2\,-\,\Lambda^2\,+\,{\vec x}^2\,-\,t^2\,\Big)
\nonumber\\
X_{n+1} &=& {\Lambda t \over z}\,,
\end{eqnarray}

\noindent where $\vec x $ has $n-1$ components and 
 $0 \le  z < \infty $. In this case the $\,AdS_{n+1}\,$ measure with 
Lorentzian signature reads

\begin{equation}
\label{metric}
ds^2=\frac {\Lambda^2 }{( z )^2}\Big( dz^2 \,+(d\vec x)^2\,
- dt^2 \,\Big)\,.
 \end{equation}

It is important to see how the compactification discussed in global coordinates 
can be realized in this system. 
The $AdS$ boundary ($\rho \,=\, \pi/2$ in global coordinates) corresponds to the 
region $\,z\,=\,0\,$ described by usual Minkowski coordinates $\vec x$ , $t$
 plus a ``point'' at infinity ($z\,\rightarrow\,\infty\,$). 
The point at infinity can not be accommodated in the original Poincare 
chart\cite{BB1,BB2} so that  we have to introduce a second coordinate system
to represent it properly.

It is convenient to introduce first an auxiliary variable that will connect the two 
charts.
Let us define the auxiliary variable $u$ as the argument of a monotonic function 
$f( u )$ such that 

\begin{equation}
\label{z}
z \,=\, {1\over f(\,u\,)}.
\end{equation}

\noindent This way the point $\,z \rightarrow \infty\,$ is mapped at 
the zero of $f( u )$. The simplest choice for this function is a linear one
as 

\begin{equation}
f (u) \,=\, c_o + c_1 u
\end{equation}

\noindent with $c_0 , c_1  $ constants. The relation (\ref{z}) is not defined for 
$u =\, -\,c_0/c_1\,$  that corresponds to the point at infinity. Also the variables 
$z$ and $u$ are not related at the point $z = 0$. As the zero value of $z$ would
be reached for infinite $u$ we can take relation  (\ref{z}) 
to be valid in the interval $\delta \le z < \infty$  for some small positive $\delta$.
This implies a finite range for $u$. For convenience we choose 
$c_0 = 1/\delta\,,\,c_1 = -1\,$ so that  $ 0 \le  u \le  1/\delta \,$ and

\begin{equation}
z \,=\, {1\over {\textstyle 1\over\textstyle\delta} \,-\,u}\,\,.
\end{equation}

\noindent Then we can define the second  coordinate chart 
$\,( z^\prime , \vec x , t )\,$  with 

\begin{equation}
\label{zprime}
z^\prime \,=\, {1\over u }\,.
\end{equation}

\noindent Now the point $z \rightarrow \infty$  is represented 
in the second chart at  $z^\prime\,=\,\delta $.
The coordinates $z$ and $z^\prime $ of the two charts are related by

\begin{eqnarray}
{1\over z^\prime} = {1\over \delta} - {1\over z} 
\end{eqnarray}

\noindent with range $\delta \le z^\prime < \infty \,$.

The metric of the second coordinate system involves a Poincare like factor

\begin{equation}
\label{metric2}
ds^2 \,=\, {\Lambda^2 \over z^{\prime\,2}} \, \Big\lbrack\,\,
{ \delta^2 \over (z^\prime - \delta )^2 } dz^{\prime\,2}\,+\,\,
{ (z^\prime - \delta )^2 \over \delta^2 } \Big( (d\vec x)^2\,
- dt^2 \,\Big)\,\Big\rbrack
\end{equation}

Now the compact $AdS$ space is described by the coordinate charts corresponding to 
eqs. (\ref{metric}) and (\ref{metric2}).  For example, for an $AdS_5$ we can calculate
the  Ricci scalar curvature for the two charts  finding

\begin{equation}
\mathit{R\ }= - 20\,{\displaystyle \frac {1}{\Lambda^{2}}} 
\end{equation}

\noindent for both, as expected since they describe parts of 
the same $AdS$ space. 

Further, with this second chart we find a horizon 
(infinite singularity in the spatial part of  $ds^2$) 
at $z^\prime = \delta$. This was not apparent in the original Poincare chart.
We are going to see in the next section that this horizon corresponds to the one
found in the  D3-brane system. Some other aspects of Poincare coordinate description
 of $AdS$ space have been studied in \cite{McI}.

\section{Branes and $AdS$ space}

The brane system is one of the Physical settings for the $AdS/CFT$ correspondence.
Let us now study the ten dimensional geometry generated by $N$ coincident D3-branes
and its relation to the compactified $AdS$ space. 
The metric can be written as\cite{HS,GKP} 
 
\begin{equation}
\label{branemetric}
ds^2 \,=\, \Big( 1 + {\Lambda^4\over r^4} \Big)^{-1/2} ( -dt^2 + d{\vec x}^2 ) +  
\Big( 1 + {\Lambda^4\over r^4} \Big)^{1/2} (dr^2 + r^2 d\Omega^2_5 )
\end{equation}

\noindent where we are using the same symbol $\Lambda$ for 
a constant that now satisfies $\Lambda^4 \,=\, N/ 2\pi^2 T_3$ where
$T_3$ is the tension of a single D3-brane. 
The metric (\ref{branemetric}) has  a horizon at $r = 0$ with zero perpendicular area 
(apart from the $S_5$ term).

It is interesting to look at the space corresponding to (\ref{branemetric}) in two 
limiting cases where it assumes simpler asymptotic forms: large and small $r$ 
compared to $\Lambda$.
Considering first the region $r >> \Lambda $ (far from the horizon) the space is 
asymptotically a ten dimensional Minkowski space:

\begin{equation}
\label{flat}
(ds^2)_{far} \,=\, -dt^2 + d{\vec x}^2  +  
dr^2 + r^2 d\Omega^2_5 
\end{equation}

\noindent Now looking at the near horizon region $r << \Lambda $ we can approximate 
the metric (\ref{branemetric}) as:

\begin{equation}
(ds^2)_{near} \,=\, {r^2 \over \Lambda^2} ( -dt^2 + d{\vec x}^2 ) +  
{\Lambda^2\over r^2}dr^2 + \Lambda^2 d\Omega^2_5 
\end{equation}

\noindent Changing the axial coordinate according to: $ z = \Lambda^2/r$,
as in ref.\cite{Malda,GKP}, 
the metric that will describe the brane system as long as
$ r/ \Lambda << 1$ takes the form

\begin{equation}
\label{metric3}
ds^2=\frac {\Lambda^2 }{ z^2}\Big( dz^2 \,+(d\vec x)^2\,
- dt^2 \,\Big)\,+\,\Lambda^2 d\Omega^2_5.
\end{equation}

\noindent corresponding to $AdS_5\times S_5\,$. 
This corresponds to the Poincar\'e chart (\ref{metric}) apart from the $S_5$ factor.
Note however that the horizon  $r=0$ which corresponds to the limit  
$z \rightarrow \infty $
is not included in this chart as a consequence of the lack of a relation between $z$ 
and $r$ at $r=0$. It is interesting to note that from the point of view of a pure $AdS$
space, one has to include this point as a 
requirement for  a consistent quantization.  Considering the  brane space this point 
is already present,  corresponding to the brane location. 
The inclusion of this point in the $AdS$ space  is possible by introducing one 
more coordinate chart as discussed in the previous section. Explicitly: the point $r=0$
corresponds to $z^\prime = \delta$. So, indeed the horizon found in the second chart at
$z^\prime = \delta$ corresponds to the brane horizon.

Let us now examine the large $r$ region of the D3-branes space. A  massless particle
moving in the $r \rightarrow \infty $ direction will arrive at an asymptotically
Minkowski space as in eq.(\ref{flat}).
Then it would spend an infinite time to reach spatial infinity.
So, the Cauchy problem is well posed for the D3-branes space
and it is geodesically complete. 
This is the physical setting of Gubser, Klebanov and Polyakov\cite{GKP}.

Further it is interesting to consider the limit $\Lambda \rightarrow \infty $
as suggested by the Maldacena conjecture. 
The larger we take $\Lambda $ the larger is the range 
of $r$ for which the $AdS$ approximation (\ref{metric3}) for the brane metric 
(\ref{branemetric})  holds. 
So one could naively disregard the asymptotic flat space region in this limit. 
Then one would find an $AdS$ space
without the boundary, where particles could enter or leave the space in finite times.
This would lead to the absence of a well defined Cauchy problem.
 
If one chooses to disregard the flat space region, boundary conditions
should be imposed at $r\rightarrow \infty $ in order to recover
physical consistency.
That means, in the limit $\Lambda \rightarrow \infty $ we should not
represent the branes space by just a pure $AdS$ space but rather by a 
compactified $AdS$ including the hypersurface at $z = 0$ besides the point
$z$ at infinity. This is the Witten's \cite{Wi} physical setting for the 
$AdS/CFT$ correspondence.

It is interesting to note that if we consider the whole space to be of the $AdS$ form
(eq.  (\ref{metric3})) there is a horizon with infinite area at  $z = 0$. 
This is not present in the D3-branes model and it is a consequence of closing the 
$AdS$ space as required for physical consistency once the asymptotic flat space region
has been removed. 
This emphasizes the differences between the spaces considered in equivalent formulations
of $AdS/CFT$ correspondence.
 
Note also that the boundary of the space defined by the metric (\ref{metric3}),
apart form the point  at $z\rightarrow \infty$, corresponds to 
$z = 0$ which naively has $8 + 1$ dimensions. 
But as $z$ approaches zero the term $ \Lambda^2 d\Omega^2_5 $ becomes irrelevant with
respect to the $AdS$ part. So we can think of the $z=0$ hypersurface as just 
$3+1$ dimensional. Then naturally the $CFT$ lives in 4 dimensions, although the 
brane model is defined in 10 dimensions.

\section{Conclusions}

The $AdS/CFT$ correspondence can be realized in different spaces.
One of them is the space generated by $N$ coincident D3-branes and another
is a compact $AdS$ space.  We have discussed  the physical consistency of 
these two formulations from the point of view of the Cauchy problem. 
The brane space is consistent thanks to the existence of a flat asymptotic region
far from the branes and of a horizon on their location.

For the compact $AdS$ space, consistency comes from the inclusion of its 
boundary:  a hypersurface at $z=0$ plus a point at $z \rightarrow \infty$. 
As this point at infinity is not  properly represented in the Poincare patch
we introduced a second coordinate chart.
In this chart that point is found at $z^\prime = \delta\,$ which corresponds to a 
horizon. This horizon was not apparent in the Poincare patch although  it is present 
in the brane system. 
This provides a nice physical interpretation for the inclusion of the point 
at infinity in the $AdS$ case.
We have also found a horizon at $z=0$ that is not present in the brane system. 
The two horizons together are responsible for the physical consistency
of the pure  $AdS$ case. 
In fact one can think that the $AdS$ boundary corresponds just to a single horizon
which confines particles inside the space.
This can be seen by looking at global coordinates where this boundary is represented 
simply as the hypersurface $\rho = \pi/2$. In this regard it is important to mention
that the connectedness of the $AdS$ boundary was proved in \cite{WY}.

\section*{Acknowledgments} 
We would like to thank Marcelo Alves for interesting discussions. 
The authors are partially supported by CNPq, FINEP and FAPERJ 
- Brazilian research agencies.



\end{document}